# A Combination 5-DOF Active Magnetic Bearing For Energy Storage Flywheels

Xiaojun Li, *Member, IEEE*, Alan Palazzolo, and Zhiyang Wang

***Abstract*— Conventional active magnetic bearing (AMB) systems use several separate radial and thrust bearings to provide a 5 degree of freedom (DOF) levitation control. This paper presents a novel combination 5-DOF active magnetic bearing (C5AMB) designed for a shaft-less, hub-less, high-strength steel energy storage flywheel (SHFES), which achieves doubled energy density compared to prior technologies. As a single device, the C5AMB provides radial, axial, and tilting levitations simultaneously. In addition, it utilizes low-cost and more available materials to replace silicon steels and laminations, which results in reduced costs and more convenient assemblies. Apart from the unique structure and the use of low magnetic grade material, other design challenges include shared flux paths, large dimensions, and relatively small air gaps. The finite element method (FEM) is too computationally intensive for early-stage analysis. An equivalent magnetic circuit method (EMCM) is developed for modeling and analysis. Nonlinear FEM is then used for detailed simulations. Both permanent magnets (PM) and electromagnetic control currents provide the weight-balancing lifting force. During the full-scale prototype testing, the C5AMB successfully levitates a 5440 kg and 2 m diameter flywheel at an air gap of 1.14 mm. Its current and position stiffnesses are verified experimentally.**

## NOMENCLATURE

| | |
|---|---|
| $R_X(i)$ | Reluctance of the i[th] {X} pole |
| $R_X^{pm}(i)$ | Reluctance of the i[th] {X} PM ring |
| $\phi_X^Y(i)$ | Flux of the i[th] {X} pole, related to {Y} |
| $F_X^Y(i)$ | MMF of the i[th] {X} pole, related to {Y} |
| $A_X(i)$ | Area of the i[th] {X} pole |
| $R_X$ | Total reluctance of the {X} pole |
| $R_X^{pm}$ | Total reluctance of the {X} magnetic ring |
| $\phi_X^Y$ | Total flux of the i[th] {X} pole, related to {Y} |
| $F_X^Y$ | Total MMF of the ith {X} pole, related to {Y} |
| $A_X$ | Total area of the {X} poles |
| $\mathbf{R}_X$ | Diagonal matrix of reluctances for {X} poles |
| $\mathbf{R}_X^{pm}$ | Diagonal matrix of reluctances for {X} PM ring |
| $\mathbf{\Phi}_X^Y$ | Flux vector of {X} pole, related to {Y} |
| $\mathbf{F}_X^Y$ | MMF vector of {X} pole, related to {Y} |
| $\mathbf{J}_n$ | $n \times n$ unit matrix |
| $\mathbf{0}_{n,m}$ | $n \times m$ zero matrix |
| $\mathbf{e}_\mathrm{n}$ | $1 \times n$ unit vector |
| $X_a$ | Subscript for the axial poles |
| $X_{ri}$ | Subscript for the {inner, outer} radial poles |
| $X_{ro}$ | Subscript for the radial poles |
| $X_r$ | Subscript for the radial poles |
| $X_t$ | Subscript for the PM poles |
| $X_{dw}$ | Subscript for the lower PM ring |
| $X_{up}$ | Subscript for the upper PM ring |
| $X^{pm}$ | Superscript for the permanent magnets |
| $X^i$ | Superscript for the control flux |
| $X^{it}$ | Superscript for the tilt control current |
| $X^{ia}$ | Superscript for the axial control current |
| $X^{ir}$ | Superscript for the radial control current |
| $z_0$ | Nominal axial air gap |
| $z_1$ | Nominal radial air gap |
| $x$ | X position of the flywheel |
| $y$ | Y position of the flywheel |
| z | Z position of the flywheel |
| $\theta_x$ | X-axis tilt motion of the flywheel |
| $\theta_y$ | Y-axis tilt motion of the flywheel |
| $(r, \psi)$ | Polar coordinate of the x-y plane |
| $g(r, \psi)$ | Tilt motion-induced airgap |
| $R^2$ | Coefficients of determination |

Manuscript accepted by IEEE. This work was supported in part by the U.S. Department of Energy and the Texas A&M Institute of Energy.

Xiaojun Li was with the Department of Mechanical Engineering, Texas A&M University, College Station, TX 77840, USA. He is now with Gotion Inc, Fremont, CA ,94538 USA (e-mail: tonylee2016@gmail.com).

Alan Palazzolo is with the Department of Mechanical Engineering, Texas A&M University, College Station, TX 77840, USA.

Zhiyang Wang was with the Department of Mechanical Engineering, Texas A&M University, College Station, TX 77840, USA. He is now with Vycon Inc, Cerritos, CA 90703 USA

| | |
|---|---|
| AMB | Active magnetic bearing |
| CAMB | Combination Active magnetic bearing |
| CRAMB | Combined radial-axial magnetic bearing |
| C5AMB | Combination 5 degree-of-freedom active magnetic bearing |
| FESS | Flywheel energy storage system |
| FEM | Finite element method |
| MMF | Magnetomotive force |
| PM | Permanent magnet |
| SHFES | Shaft-less, hub-less, high-strength steel energy storage flywheel |

# I. Introduction

ACTIVE Magnetic Bearings have many advantages over conventional bearings. They require minimal maintenance and have fewer environmental impacts by eliminating the lubrication systems indispensable for fluid-film bearings. In addition, compared to rolling element bearings, they offer no friction loss and higher operating speed [1] due to magnetic levitation's non-contact nature. As a result, magnetic bearings have been increasingly used in industrial applications such as compressors, pumps, turbine generators, and flywheel energy storage systems (FESS) [2].

Magnetic bearing supported rotating machinery, whether based on a vertical or horizontal rotor, needs several subsystems responsible for the radial and axial levitations. Typically, two radial and one axial AMB are used for a 5-DOF levitation [3]. Combination magnetic bearings simplify the system, reduce cost, and improve rotor dynamics and control performances by eliminating extra actuators. Several combination designs have been proposed with different focuses. Among one of the early works, [4] presents the magnetic bearing system for a 42,000 RPM flywheel. The system combines one radial bearing with the axial bearing, reducing the number of units from three to two. Another early work [5] presents a 500,000 rpm, combined radial-axial magnetic bearing (CRAMB) system for a 1kw PM machine, in which two bearing units are used. The proposed magnetic bearing system in [6] includes one passive magnetic bearing and one hybrid radial magnetic bearing. The system is designed for a control moment gyroscope flywheel used in agile satellites. In [7], a PM-biased axial hybrid magnetic bearing is presented. It has four-segment poles to control 3-DOF. In [8], the authors propose a 4-DOF magnetic bearing capable of the radial and tilting controls.

More recently, [9] discusses the design methodology for CRAMBs. An asymmetric factor is proposed to facilitate the design process. In [10], a three-pole CRAMB is presented. Compared to side-by-side CRAMBs, the benefits include fewer electronic switches and the shortened bearing shaft length. Based on our survey, in most CRAMB designs, multiple magnetic bearings are still needed to provide the 5-DOF levitation. A typical side-by-side CRAMB uses a small extension on the shaft as the axial flux path. They are more suitable for applications with light or horizontally placed shafts but less for supporting heavy flywheels. CRAMBs also have couplings between axes due to shared bias fluxes, putting some constraints on the design, such as the maximum axial and radial bearing capacities [9], [11]. Apart from combination designs, conical magnetic bearings can also provide radial and axial levitations together. The conical AMB has a small radial extent, making it suitable for smaller devices with limited spaces [12].

Texas A&M University has built a novel AMB-supported 100KWh Shaft-less, high-strength steel FESS that features an innovative shaft-less flywheel [13]–[15]. The flywheel, which weighs 5540 kg, is constructed of high-strength steel (AISI 4340). Its outer diameter is 2133 mm, and its height is 203 mm. The FESS has a designed energy capacity of 100KWh, powered by a 100 KW coreless permanent-magnet-synchronous motor/generator (PMSM) [16]. This paper presents the modeling, analysis, and validation of a novel PM-biased combination magnetic bearing for the shaft-less flywheel. Challenges in designing and building such an integrated and large-scale magnetic bearing include 1) *Due to the shared flux paths between radial, tilting, and axial poles, there are coupling effects between magnetic poles*. 2) *Due to the shared flux paths, it is impossible to design for each axis separately* 3) *Due to its large size, geometry complexity, and complex airgaps, the magnetic bearing is difficult to simulate by FEM, leading to an extended design cycle.* 4) *Due to the large size and the use of low-cost materials, there is less tolerance for modeling errors.*

To address these challenges, we created an EMCM model that incorporates 5-DOF motions and current excitations altogether, whereas previous EMCMs usually treat them independently [17]. For simplicity, flux leakages and fringing effects are not considered. The equivalent circuit utilizes a lumped element approach. It is applied during the initial design phase to study the parameters and coupling effects quickly but efficiently. In the later stage, a nonlinear FEM is used for fine-tuning and validating the parameters. The C5AMB's design also includes unique features. For example, each axis' coefficients and load capacity can be designed independently and adjusted on-site to compensate for modeling errors. Finally, the C5AMB's current and position stiffnesses are obtained experimentally during the magnetic levitation [18]. This paper's contributions include: 1) *A single CAMB device replaces several magnet bearings to support a 5400 kg flywheel reliably*. 2) *A novel CAMB design in which each axis's stiffness can be designed and adjusted individually.* 3) *An effective method of designing integrated, large-scale magnetic bearing systems*.

The remainder of this paper is organized as follows: An overview of the SHFES is given in section II. The design, working principle, and modeling of the C5AMB are presented in section III. In section IV, an EMCM is used to analyze the magnetic bearing's capacity and stiffness coefficients. The results are verified with FEM simulations. The coupling effects between magnetic poles are also discussed in this section. Finally, section V gives the measurements of position and current stiffness of the fully assembled C5AMB-flywheel system.

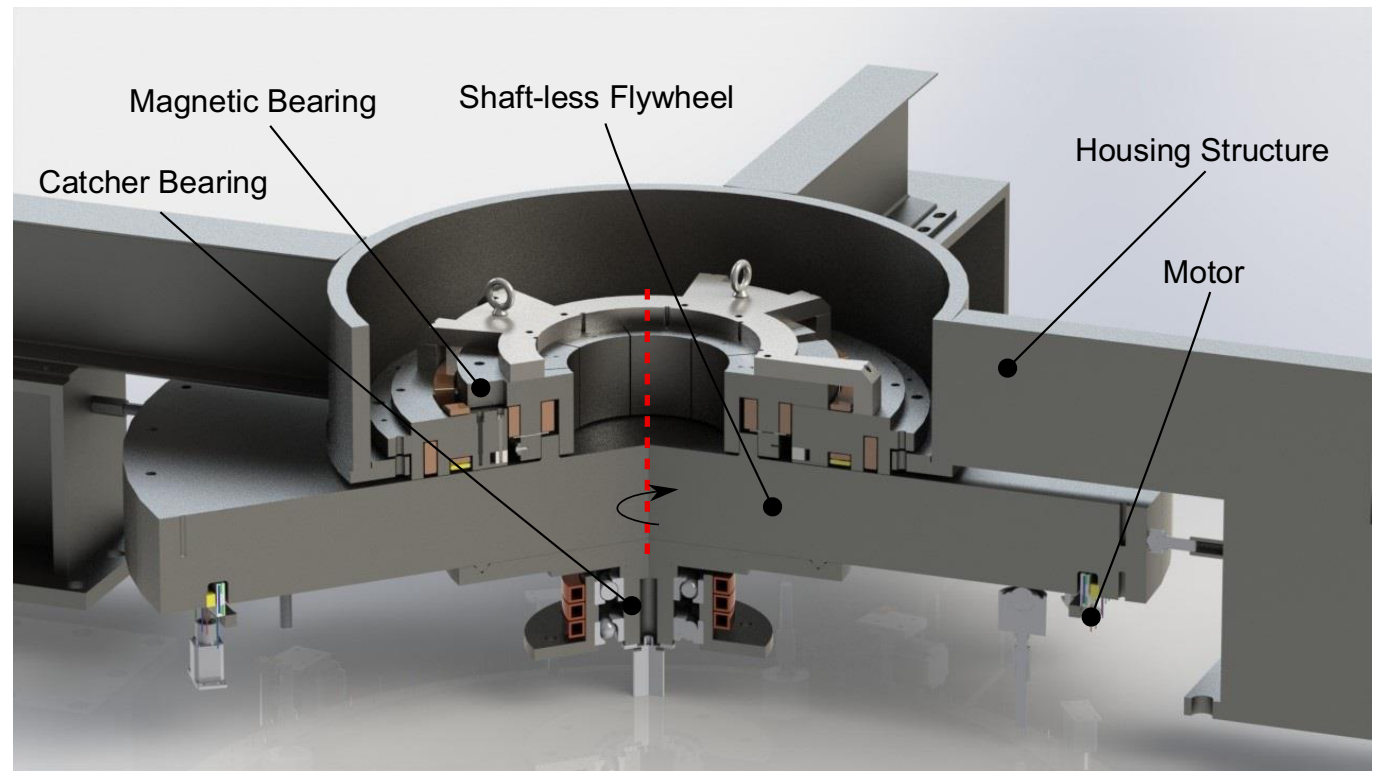

Fig. 1. The C5AMB-SHFES assembly. The C5AMB is a combination, PM-biased, homopolar, and active magnetic bearing capable of providing 5-DOF levitation. The flywheel's rotational axis is marked with dashed red line.

## II. Overview of the SHFES

A flywheel stores energy in the form of rotational kinetic energy, which is given by

$$E = \frac{1}{2} I \omega^2 \tag{1}$$

where $I$ denotes the moment of inertia and $\omega$ denotes the spin speed. For higher energy capacity, flywheels are designed to achieve high spin speed, leading to failures when inertia loads develop excessive stress. As such, the energy capacity of a flywheel is bounded by its material's yield stress. Therefore, it is desirable to design the flywheel to have lower and evenly distributed stress to avoid failure caused by stress concentrations. When spinning, a flywheel is subject to both hoop ($\sigma_\theta$) and radial ($\sigma_r$) stress, the distribution of which is greatly influenced by its geometric profile.

The proposed FESS removes the shaft and borehole to improve the stress distribution [14], [19]. As a result, its energy density is doubled to conventional FESSs that consist of a borehole rotor and a shrink-fitted shaft. High-strength steel is adopted as the building material so that the flywheel can be forged as a solid disc. Compared to prior steel flywheel designs (3.5~8.3 Wh/kg [20]), the shaft-less flywheel's specific energy is doubled to 18.2 Wh/kg [19]. Composite flywheels have achieved higher specific energy (50-100 Wh/kg) when only considering the rotor [21], [22]. However, the composite rotor only takes a small portion of the entire system weight. When considering the whole flywheel, one of the reported composite designs [23] reached 11.7 Wh/kg. This value is halved when including the auxiliary components. The reported energy density [23] is 25 kWh/m$^3$, whereas the SHFES has an energy density of 35 kWh/m$^3$. In conclusion, the SHFES provides competitive specific energy (energy per mass) and energy density (energy per volume) to composite flywheels at a lower cost.

As depicted in Fig. 1, the C5AMB, motor, catcher bearing, and the housing structure are designed to be integrated with the shaftless flywheel, giving the SHFES a high integration level. A high degree of integration allows maximizing energy density [24] and easier adaption to industrial applications. The SHFES is mainly targeted for large-scale utility applications but can be easily adapted to smaller-scale and broader applications. Particularly, flywheels have a high potential in fast charging for electric vehicles. Using energy storage devices for fast charging reduces the cost of infrastructure upgrades. Compared to other energy storage technologies like li-ion batteries, flywheels have longer life cycles and higher power density. Other advantages include operability under low/high temperatures, accurate state-of-charge, and recyclability [25].

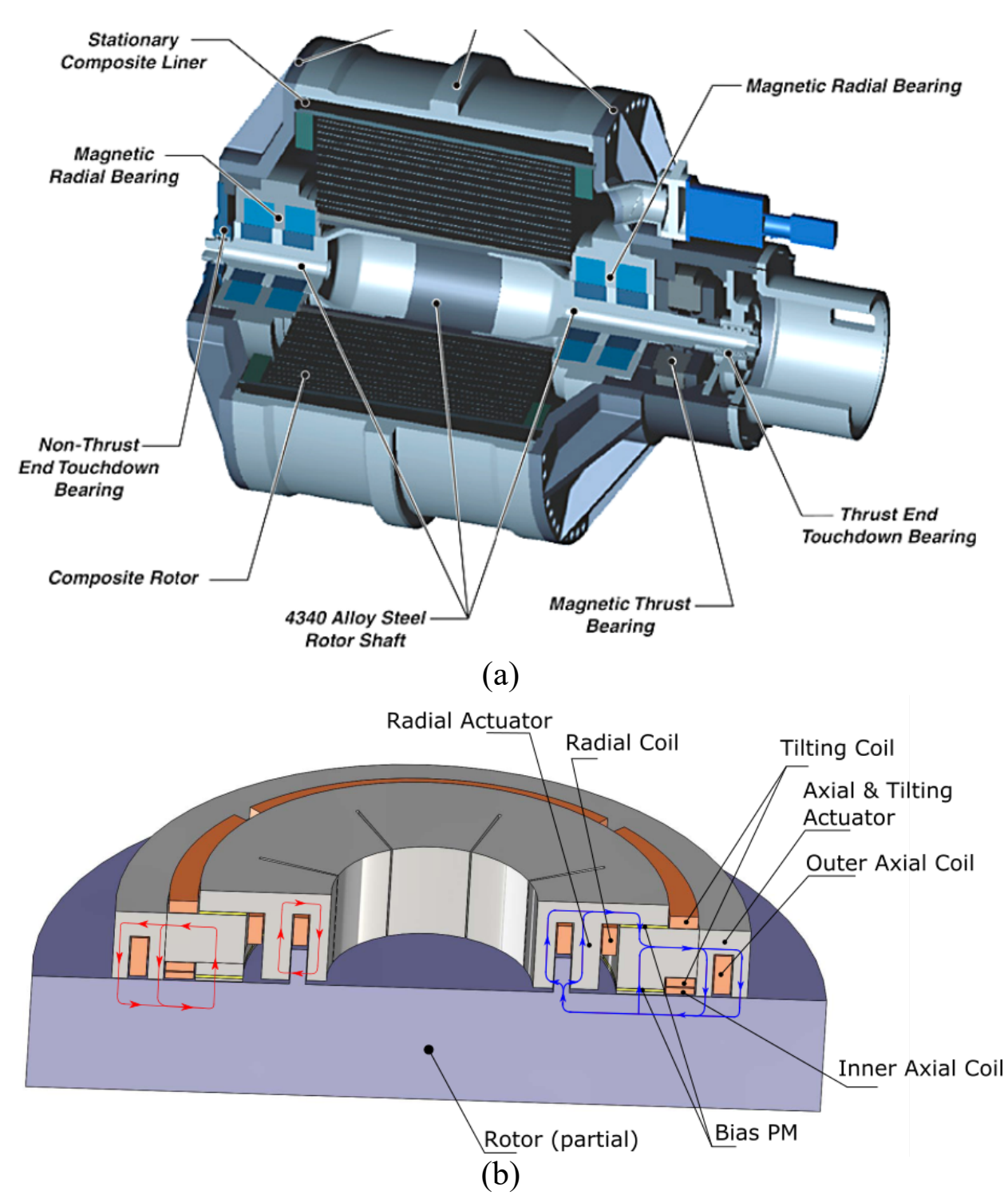

Fig. 2. (a) A typical magnetic bearing system [23] includes a long shaft and several distributed components to provide 5-DOF levitation. (b) Close-up view of the combination magnetic bearing structure [14], with key components labeled. The C5AMB is highly integrated with the flywheel. The blue and red lines represent bias fluxes and control fluxes, respectively.

TABLE I
Design Parameters of the C5AMB-SHFES

| | SHFES | C5AMB |
|---|---|---|
| Nominal air gaps | - | 1.14 mm |
| Outer diameter (mm) | 2133 | 1106 |
| Height (mm) | 203 | 165 |
| Weight (kg) | 5540 | 544 |
| Material (AISI) | 4340 | 1018 |
| Relative permeability | 200 | 1000 |
| Permanent magnet grade | - | N48 |
| Saturation flux density (T) | 0.8 | 1.5 |

## III. Working Principle and Modeling of the Combination Magnetic Bearing

### A. Design and working principles

Since the new flywheel design has eliminated the shaft, traditional AMBs, such as the one depicted in Fig. 2a, do not apply to the SHFES. A C5AMB is introduced to both accommodate the shaft-less design and reduce the system complexity and costs. The C5AMB includes two major parts that are assembled mechanically and spaced by the permanent magnets. The first component is the radial actuator. As depicted in Fig. 2b, it is an inner circular part with radial poles attached circumferentially. There are eight radial sub-actuators, each of which includes an inner and outer radial pole. Under the operation mode, the rotor ring is inserted between radial pole pairs to complete the flux path. There are eight radial coils mounted on the outer poles to provide control fluxes. As Fig. 2b shows, the other part is a circular component that harbors the axial and PM poles. It includes four ring-shaped structures for the PM pole and axial poles. Four tilting coil windings occupy each of the PM/axial poles. Two sizeable circular coil windings with diameters of 0.8 m and 0.9 m are used for providing the axial control flux throughout all PM/axial poles.

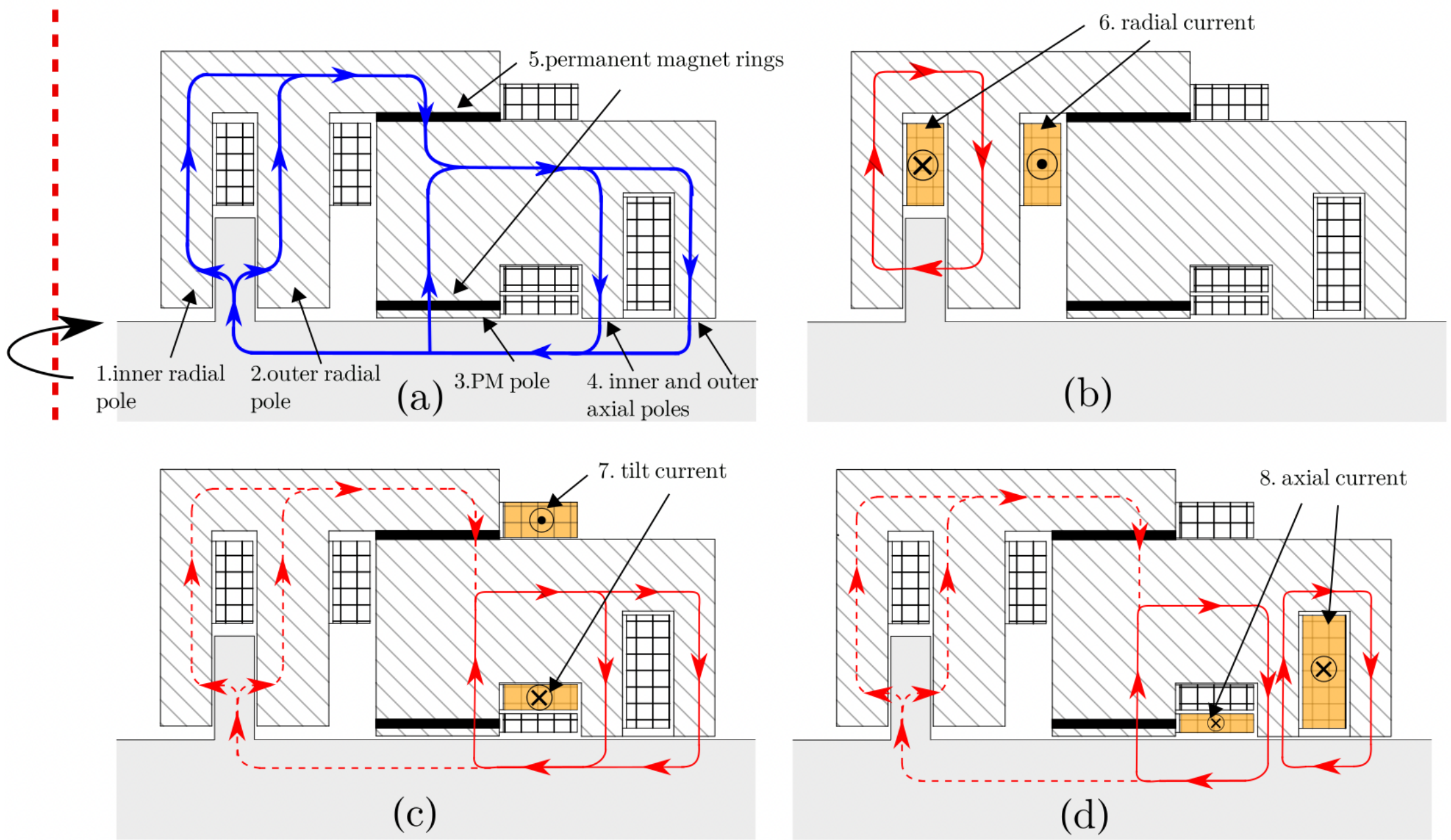

Fig. 3. The axial-symmetrical section view of the C5AMB-flywheel and their working modes. The flywheel's rotational axis is depicted as a dashed line. PM and currents generated fluxes are highlighted in blue and red, respectively. Each of the subplots is for illustrating: (a) PM flux (b) radial control flux (c) tilting control flux (d) axial control flux. The components are labeled as the follows: 1- inner radial pole, 2- outer radial pole, 3- PM pole, 4- inner axial pole and outer axial pole, 5- permanent magnets, 6- radial coil and current, 7- tilt coil and current, 8- axial coil and current.

The magnetic bearing is permanent-magnet-biased to offset the flywheel's weight. The tilting and axial actuators share the PM bias fluxes supplied by two sets of PM ring sections. This particular PM configuration enables a homopolar design, which alleviates losses on the rotor. If the radial poles of the AMB are not biased (heteropolar), magnetic fields with reversed polarities are needed to generate a net radial force on a specific axis. As a result, when the flywheel rotates, it will travel through magnetic fields of different polarities, which induces more significant eddy currents and hysteresis losses. With a homopolar AMB, the flywheel suffers less loss by traveling through magnetic fields with the same polarity and less flux density differences. The homopolar design also can be achieved using a DC current. Nevertheless, the current consumes extra energy and reduces the power amplifiers' effective control current capacity.

One PM ring is installed between the radial and axial/tilting part to serves as a joint. The other one is installed at the PM poles. Two sets of PMs are used instead of one for the following considerations:

1) Two PMs provide adequate axial bias flux densities (around 0.8T) to support the 5540kg flywheel and decent radial bias flux densities (0.5T to 0.6T).
2) Bias flux for different axes can be modified by adjusting the two magnet rings (Details are covered in B.1)).
3) Bias flux can be evenly distributed without local saturations that a single dominant PM ring can cause.
4) Magnets have a high reluctance to act as barriers between the axial and radial flux paths to reduce flux leakage.

The AMB's components and various working modes are further illustrated in Fig. 3. As shown in Fig. 3a, The two PM rings are placed such that their magnetic poles oppose each other. The top and bottom magnets use the S-N and N-S orientation, respectively. The bias flux for radial, tilt, and axial poles, are colored in blue. For the inner part of the C5AMB, which provides radial levitation, the bias flux travels through the flywheel's radial ring and evenly diverges to the inner and outer radial poles. For the outer part, it travels through the axial poles and returns through the PM poles. Notice that the rotational axis is for illustrating only. Its physical location is further to the left. As depicted in Fig. 3b, a net radial force is generated by strengthening the flux in one of the radial pole pairs and weakening the opposite one. As shown in Fig. 3c, the moment control is realized by applying control currents to a subset of the axial and PM poles to create flux variation with respect to the moment arm, which will results in a net moment but no axial force. The axial control force is generated by applying currents to the two large circular coil windings, As depicted in Fig. 3d.

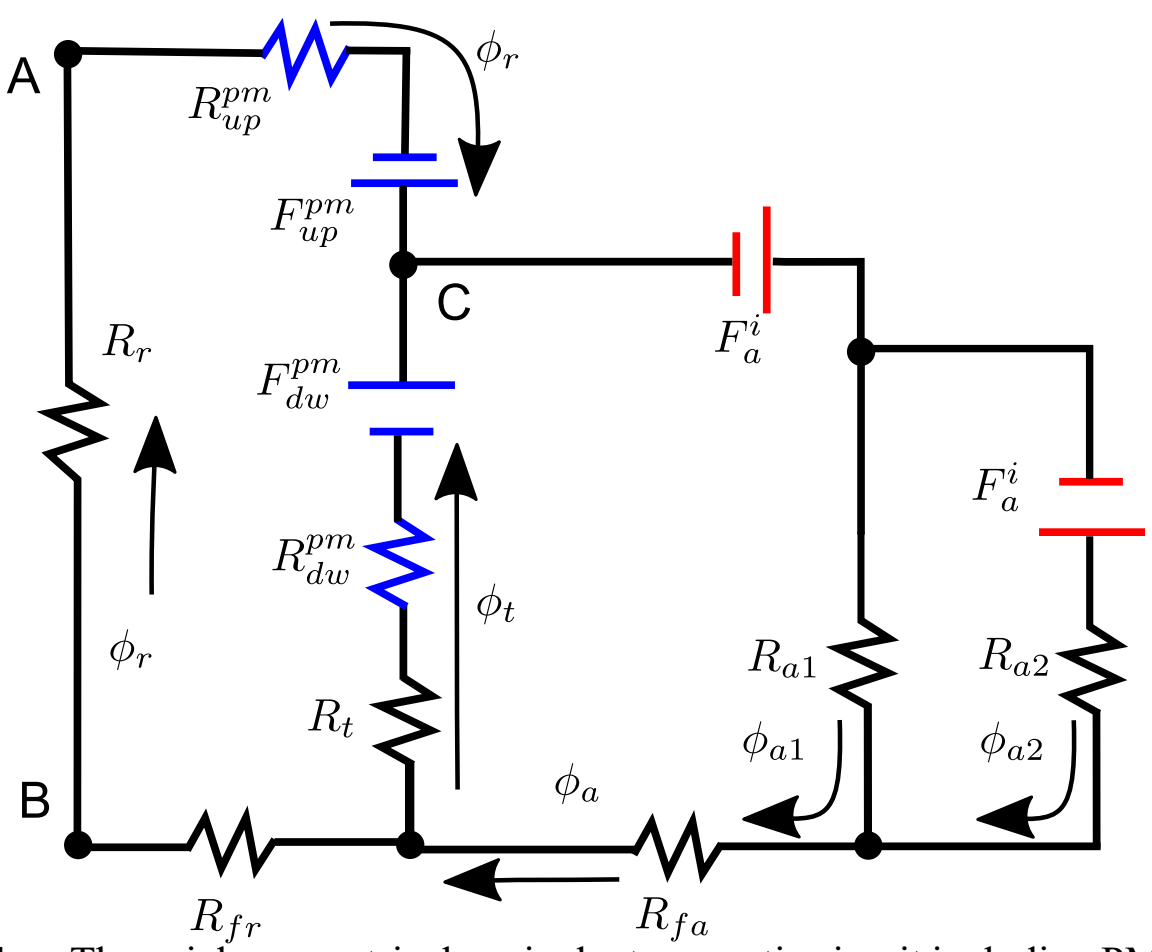

Fig. 4. The axial-symmetrical equivalent magnetic circuit including PM and axial coil MMFs.

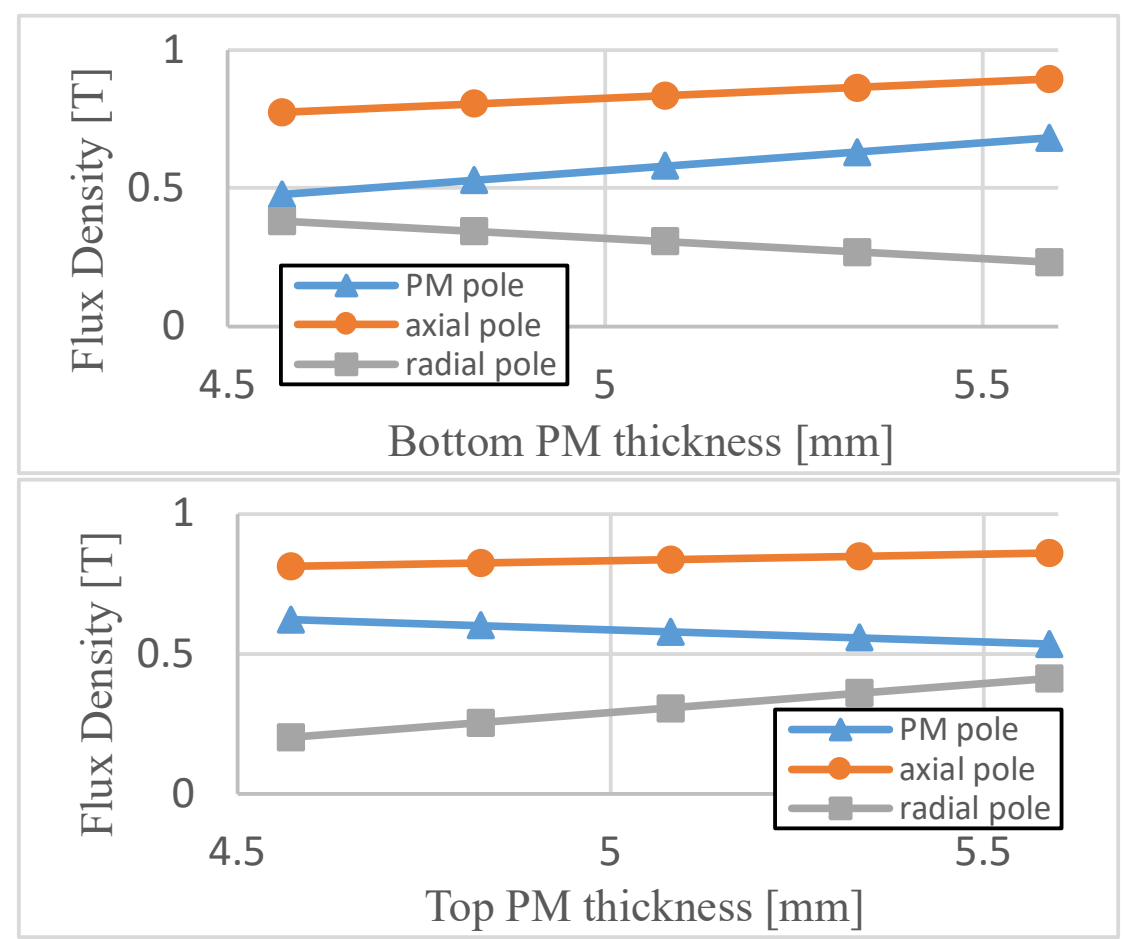

Fig. 5. The average bias flux density of magnetic poles with respect to the thickness of permanent magnets.

## *B. Modeling of the C5AMB*

In this section, we propose an equivalent circuit model for the C5AMB. The model is used in the initial design and analysis. In subsection 1), a simple 1D model is used to calculate the bias flux. The analysis shows that, like in a conventional distributed magnetic bearing system, bias flux density for each axis of the C5AMB can be adjusted independently. Next, in subsection 2), we derive the flux changes caused by the flywheel's translational and rotational motions. Then, subsections 3) to 5) present the models for calculating the axial, tilt, and radial control fluxes. Finally, subsection 6) derives the magnetic force and moment.

### *1) PM bias flux model*

As illustrated in Fig. 4, a simple axial symmetric circuit is used for modeling the bias flux distribution. This model is also used for deriving the axial control fluxes and serves as a foundation for the later analysis. Air gap fluxes are noted as follows: the flux at PM pole air gap ($\phi_t$), the inner axial pole ($\phi_{a1}$), the outer axial pole ($\phi_{a2}$), the inner radial ($\phi_{ri}$) and the outer radial pole ($\phi_{ro}$). The combined fluxes of axial and radial poles are denoted as $\phi_a$ and $\phi_r$. The bias fluxes' solution is given by

$$
\begin{aligned}
\phi_a^{pm} &= \frac{1}{L}\left[\gamma F_{dw}^{pm} + \beta F_{up}^{pm}\right] \\
\phi_t^{pm} &= \frac{1}{L}\left[\alpha\left(F_{dw}^{pm} - F_{up}^{pm}\right) + \gamma F_{dw}^{pm}\right] \\
\phi_r^{pm} &= \frac{1}{L}\left[\alpha\left(F_{up}^{pm} - F_{dw}^{pm}\right) + \beta F_{up}^{pm}\right]
\end{aligned} \tag{2}
$$

where $F_{up}^{pm} = H_c l_{up}$ and $F_{dw}^{pm} = H_c l_{dw}$ are the total MMF of the top and bottom permanent magnets. The reluctance variables in (2) are given by

$$
\begin{aligned}
L &= \alpha\beta + \beta\gamma + \gamma\alpha \\
\alpha &= R_a + R_{fa} \\
\beta &= R_t + R_{dw}^{pm} \\
\gamma &= R_r + R_{up}^{pm} + R_{fr}
\end{aligned} \tag{3}
$$

where $R_a, R_t$ and $R_r$ are the equivalent magnetic reluctance of axial, PM, and radial poles. Since the flywheel is made of high-strength steel with low relative permeability (200), its reluctance is included in the model. The equivalent reluctances from axial to PM poles ($R_{fa}$) and from PM to radial poles ($R_{fr}$) are estimated by the dimension first and verified by FEM. They are not dependent on the air gaps and invariant to the flywheel's motions. The magnetic bearing's reluctance is ignored due to the high relative permeability (1000).

According to (2), increasing the top MMF ($F_{up}^{pm}$) or decreasing the bottom MMF ($F_{dw}^{pm}$) will lead to a more substantial radial pole bias flux ($\phi_r^{pm}$) but a weaker PM pole bias flux ($\phi_t^{pm}$). The axial pole bias flux ($\phi_a^{pm}$) benefits from both top and bottom MMFs. In summary, the effects of different permanent magnet thicknesses on bias flux densities are depicted in Fig. 5. Bias flux densities directly influence the actuator's position and current stiffness, as well as load capacity. Therefore, the magnetic bearing's characteristics for each axis can be adjusted by choosing different thicknesses or materials for the two PM rings. The C5AMB's design also allows changing the magnets by simply lifting the radial actuator, making it easy to adjust the load capacity and stiffnesses for on-site. Notice that the radial fluxes between inner and outer poles are equal only when they have the same reluctance, which is often hard to achieve. Nevertheless, the net radial force is close to zero as long as the rotor is placed at the center.

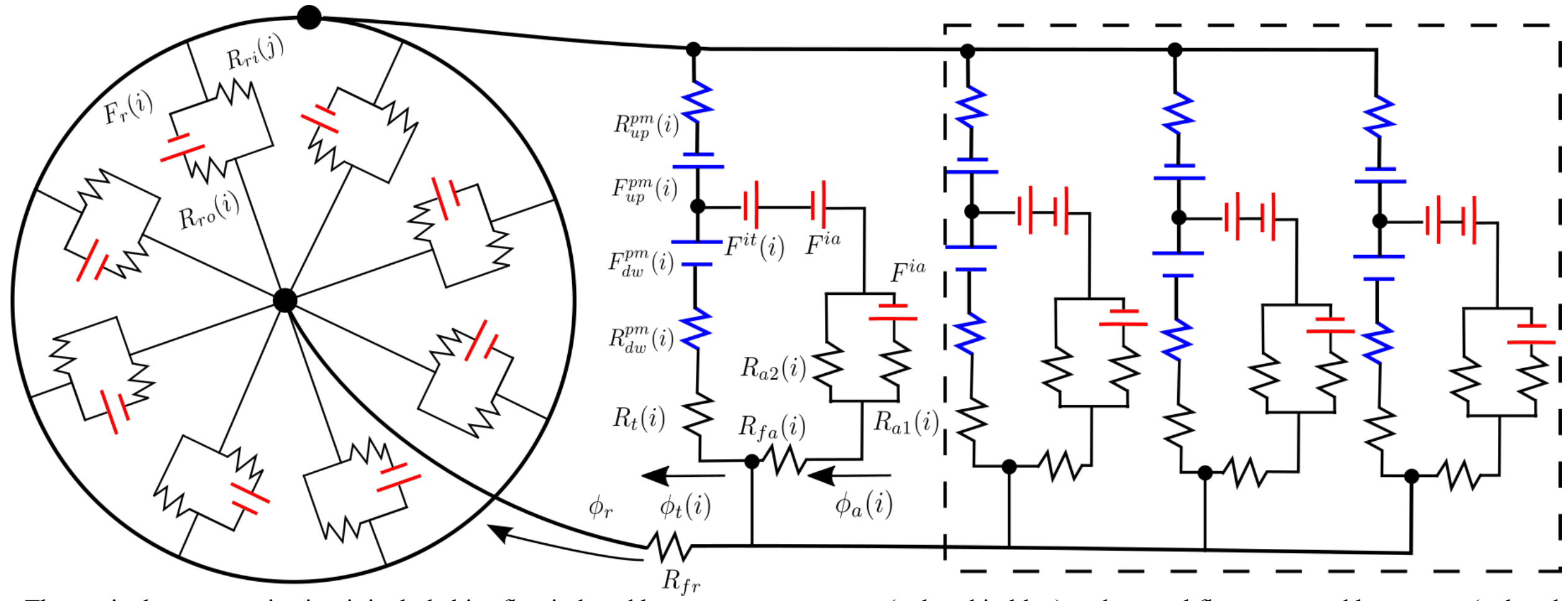

Fig. 6. The equivalent magnetic circuit include bias flux induced by permanent magnets (colored in blue) and control flux generated by current (colored in red). The left circular part is the radial magnetic structure with 8 pairs and 16 poles in total. The right part models the four sections for axial and tilt axes.

*2) Fluxes caused by the flywheel's motions*

As depicted in Fig. 6, the 3D equivalent circuit model consists of four axial and PM poles and eight radial poles, whose reluctances are functions of the air gaps that are determined by the flywheel's translational and rotational positions. The radial ($R_r(i)$), PM ($R_t(i)$) and axial ($R_{a1}(i)$, $R_{a2}(i)$) reluctances are given by

$$\begin{aligned} R_r(j) &= \frac{z_1 + g_r(j)}{\mu_0 A_r(j)} \\ R_t(i) &= \frac{1}{\iint \frac{\mu_0}{z_0 + g_t(i)} r dr d\psi} \\ R_{a1}(i) &= \frac{1}{\iint \frac{\mu_0}{z_0 + g_{a1}(i)} r dr d\psi} \\ R_{a2}(i) &= \frac{1}{\iint \frac{\mu_0}{z_0 + g_{a2}(i)} r dr d\psi} \end{aligned} \tag{4}$$

where $z_0$ is the nominal axial air gap, $z_1$ is the nominal radial air gap. $i = 1, \ldots, 4$ and $j = 1, \ldots, 8$ are the enumerations for the axial quadrants and radial pole pairs. Since the air gap $g(i)$ varies along the pole surface, the reluctances of axial and PM poles are derived using double integral in the polar coordinate $(r, \psi)$. The axial air gap changes, $g_t(i)$, $g_{a1}(i)$ and $g_{a2}(i)$ are caused by the translational and attitude motions of the flywheel, given by

$$g(r, \psi) = r[\sin\theta_x \sin\psi - \sin\theta_y \cos\psi] + z \tag{5}$$

where $\theta_x, \theta_y$ are the x-axis and y-axis tilting angles of the flywheel, $z$ is the flywheel's vertical displacement. The integral intervals for $(\psi, r)$ depends on each pole's location and size. The radial air gaps $g_r(j)$ for each pole is determined by the flywheel's radial position (x, y) relative to the AMB, defined as

$$\mathbf{g}_r \approx \begin{bmatrix} +.38x - .92y & -.38x - .92y \\ -.92x + .38y & -.92x - .38y \\ -.38x + .92y & +.38x + .92y \\ +.92x - .38y & +.92x + .38y \end{bmatrix}^T . \tag{6}$$

With the reluctances and air gaps defined, the fluxes caused by the flywheel's motion is governed by

$$\begin{aligned} &\phi_a^{pm}(i)\alpha(i) + \phi_t^{pm}(i)\beta(i) = F_{dw}^{pm}(i) \\ &(R_r + R_{fr})\phi_r^{pm}[\phi_a^{pm}(i) - \phi_t^{pm}(i)]R_{up}^{pm}(i) \\ &\qquad +\phi_a^{pm}(i)\alpha(i) = F_{up}^{pm}(i) \end{aligned} \tag{7}$$

the solution of which can be summarized in a vector form,

$$\begin{aligned} \begin{bmatrix} \mathbf{\Phi}_a \\ \mathbf{\Phi}_t \end{bmatrix} &= \begin{bmatrix} \boldsymbol{\alpha} & \boldsymbol{\beta} \\ \boldsymbol{\alpha} + M_1 & -M_1 \end{bmatrix}^{-1} \begin{bmatrix} \mathbf{F}_{dw}^{pm} \\ \mathbf{F}_{up}^{pm} \end{bmatrix} \\ \phi_r^{pm} &= \mathbf{e}_4[\mathit{\Phi}_a - \mathit{\Phi}_t] \\ M_1 &= \mathbf{R}_{up}^{pm} + \mathbf{J}_4(R_r + R_{fr}) \end{aligned} \tag{8}$$

where $[\mathbf{\Phi}_a \quad \mathbf{\Phi}_t]^T$ is the axial and tilt flux vector, $[\mathbf{F}_{dw}^{pm} \quad \mathbf{F}_{up}^{pm}]^T$ is the PM-MMF vector, $\mathbf{e}_n$ is a $1 \times n$ unit vector, and $\mathbf{J}_n$ is an $n \times n$ unit matrix. $\boldsymbol{\alpha}$ and $\boldsymbol{\beta}$ are the diagonal matrices for axial and PM pole reluctance from (3). From the combined radial flux $\phi_r^{pm}$, the individual flux for each radial pole is derived by $\phi_r(j) = \phi_r \times R_r / R_r(j)$, where $R_r$ is the combined reluctance of radial poles.

*3) Axial Control flux*

The model from Fig. 4 is also used to derive the axial control flux. For the two axial coils, they are designed to have the same turns $N_a$ and current $i_a$. The axial control flux be calculated as

$$\phi_{a1}^{ia} = \frac{F^{ia}(R_{a2} - \beta||\gamma - R_{fa})}{(\beta||\gamma + R_{fa})R_{a1} + R_{a2}R_{a1} + R_{a2}(\beta||\gamma + R_{fa})}$$

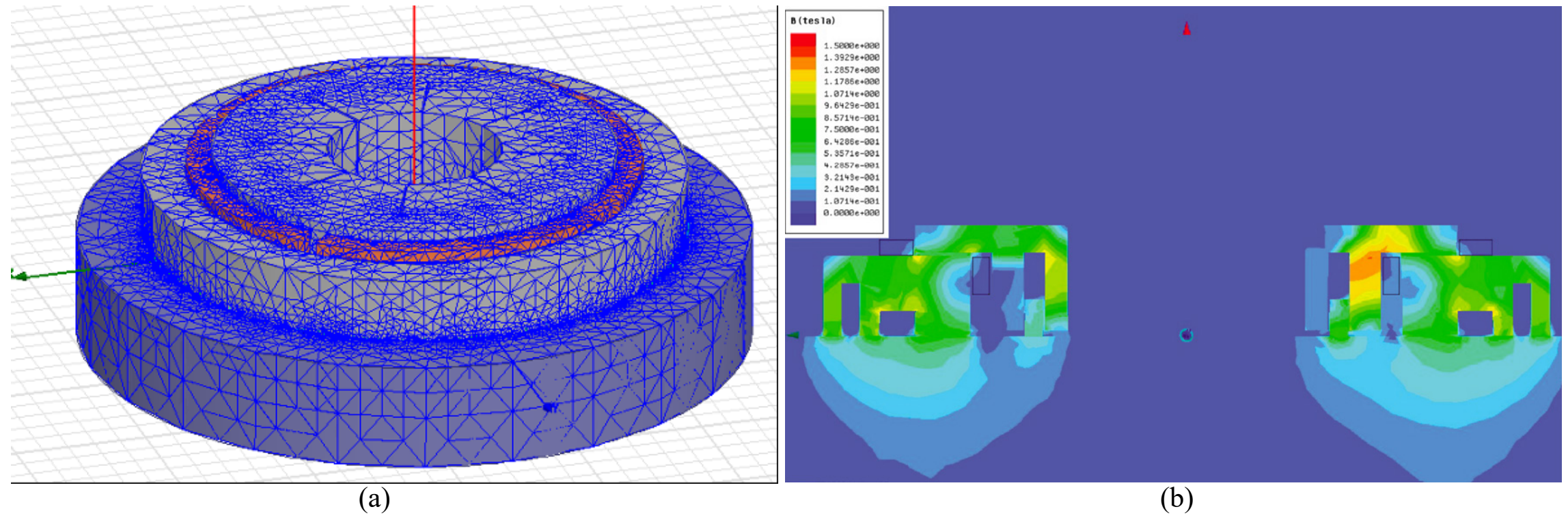

(a) (b)

Fig. 7. (a) The 3D, solid mesh FEM model of the C5AMB-SHFES. (b) A section view of the flux density, with 700 AT radial current applied

$$\phi_{a2}^{ia} = \frac{F^{ia}(2R_{a1} + \beta||\gamma + R_{fa})}{(\beta||\gamma + R_{fa})R_{a1} + R_{a2}R_{a1} + R_{a2}(\beta||\gamma + R_{fa})}$$

$$\phi_{t}^{ia} = (\phi_{a1}^{ia} + \phi_{a2}^{ia})\frac{(\beta||\gamma)}{\beta} \tag{9}$$

where $F^{ia} = N_a i_a$ is the MMF created by axial control current.

#### 4) *Tilting control flux*

The tilting control circuit is similar to the PM circuit defined in (8), with the PM's MMFs replaced by the MMFs generated by tilt currents:

$$\begin{bmatrix} \mathbf{\Phi}_a^{it} \\ \mathbf{\Phi}_t^{it} \end{bmatrix} = \begin{bmatrix} \boldsymbol{\alpha} & \boldsymbol{\beta} \\ \boldsymbol{\alpha} + M_1 & -M_1 \end{bmatrix}^{-1} \begin{bmatrix} \mathbf{F}^{it} \\ \mathbf{F}^{it} \end{bmatrix} \tag{10}$$

where $\mathbf{F}^{it}$ is the MMF vector created by the tilting currents.

#### 5) *Radial control flux*

The radial portion of the magnetic bearing has eight pole-pairs in total. Allocation of the radial flux is determined by the flywheel's radial position and the radial control currents. The governing equation is given by

$$\begin{aligned} \phi_{ri}^{ir}(j)R_{ri}(j) - \phi_{ro}^{ir}(j)R_{ro}(j) - F^{ir}(j) = 0 \\ \phi_{r}^{ir}R_{nr} + \phi_{ro}^{ir}(j)R_{ro}(j) = 0 \end{aligned} \tag{11}$$

where $R_{nr}$ is the total non-radial reluctance, $\phi_r^{ir} = \sum_{j=1}^{8}[\phi_{ri}^{ir}(j) + \phi_{ro}^{ir}(j)]$ is the total radial control flux. The solution for (11) is given by a set of 16 by 16 linear equations, as follows

$$\begin{bmatrix} \mathbf{\Phi}_{ri}^{ir} \\ \mathbf{\Phi}_{ro}^{ir} \end{bmatrix} = \begin{bmatrix} \mathbf{R}_{ri} & -\mathbf{R}_{ro} \\ \mathbf{J}_8(R_{nr}) + \mathbf{R}_{ri} & \mathbf{J}_8(R_{nr}) \end{bmatrix}^{-1} \begin{bmatrix} \mathbf{F}^{ir} \\ \mathbf{0}_{8,1} \end{bmatrix} \tag{12}$$

where $\mathbf{R}_{ri}$ and $\mathbf{R}_{ro}$ are the diagonal matrix of inner and outer radial pole reluctance. $\mathbf{F}^{ir}$ is the MMF vector contributed by radial control currents.

#### 6) *Force and moment*

Based on the superposition principle, the fluxes for all magnetic poles are given by

$$\begin{aligned} \phi_{a1}(i) &= \phi_{a1}^{pm}(i) + \phi_{a1}^{it}(i) + \phi_{a1}^{ia}(i) \\ \phi_{a2}(i) &= \phi_{a2}^{pm}(i) + \phi_{a2}^{it}(i) + \phi_{a2}^{ia}(i) \\ \phi_{t}(i) &= \phi_{t}^{pm}(i) + \phi_{t}^{it}(i) + \phi_{t}^{ia}(i). \\ \phi_{ri}(j) &= \phi_{ri}^{pm}(j) + \phi_{ri}^{ir}(j) \\ \phi_{ro}(j) &= \phi_{ro}^{pm}(j) + \phi_{ro}^{ir}(j) \end{aligned} \tag{13}$$

The magnetic force of the $j^{th}$ radial pole is given by

$$\begin{aligned} f_r(j) &= \int lr\left(\frac{1}{2\mu_0}B_g^2(j)\right)\cos\theta d\theta \\ &\approx 0.765\frac{A_r}{2\mu_0}B_g^2(j) \end{aligned} \tag{14}$$

where $B_g(j)$ denotes the flux density and $\mu_0$ represents the permeability of the free space. The magnetic force $f(i)$ of the $i^{th}$ axial/PM pole is given by

$$f_a(i) = \iint\left(\frac{\mu_0 F^2(i)}{2(z_0 + g(i))^2}\right)rdrd\psi \tag{15}$$

where $F(i)$ is the combined MMF of the $i^{th}$ axial/PM pole. The air gap $g(i)$ is defined in (5). The magnetic moment $M(i)$ of the $i^{th}$ axial/PM pole is given by

$$M(i) = \iint\left(\frac{\mu_0 F^2(i)}{2(z_0 + g(i))^2}r\cos\psi\right)rdrd\psi. \tag{16}$$

## IV. Analysis of the C5AMB

With the equivalent circuit fully developed, we now consider the initial design. Design parameters of the C5AMB include: 1) The nominal air gap of each magnetic pole. 2) The surface area of each magnetic pole. 3) The thickness and magnetic grade of the PM rings. 4) coil turns. The parameters are selected to achieve the following design targets: 1) The magnetic bearing can provide the weight-lifting force at the nominal air gap. The targeted air gap of 1.14 mm is based on several design factors, including the overall dimension, machining accuracy, the linear range of the BH curve, and the position sensor's measurement range. 2) All magnetic poles have the bias flux density (0.5T~0.8T) close to half of the saturation point (1.5T), which gives the control flux the most freedom. 3) Coil turns are calculated from the MMFs and the max current rating of the power amplifiers.

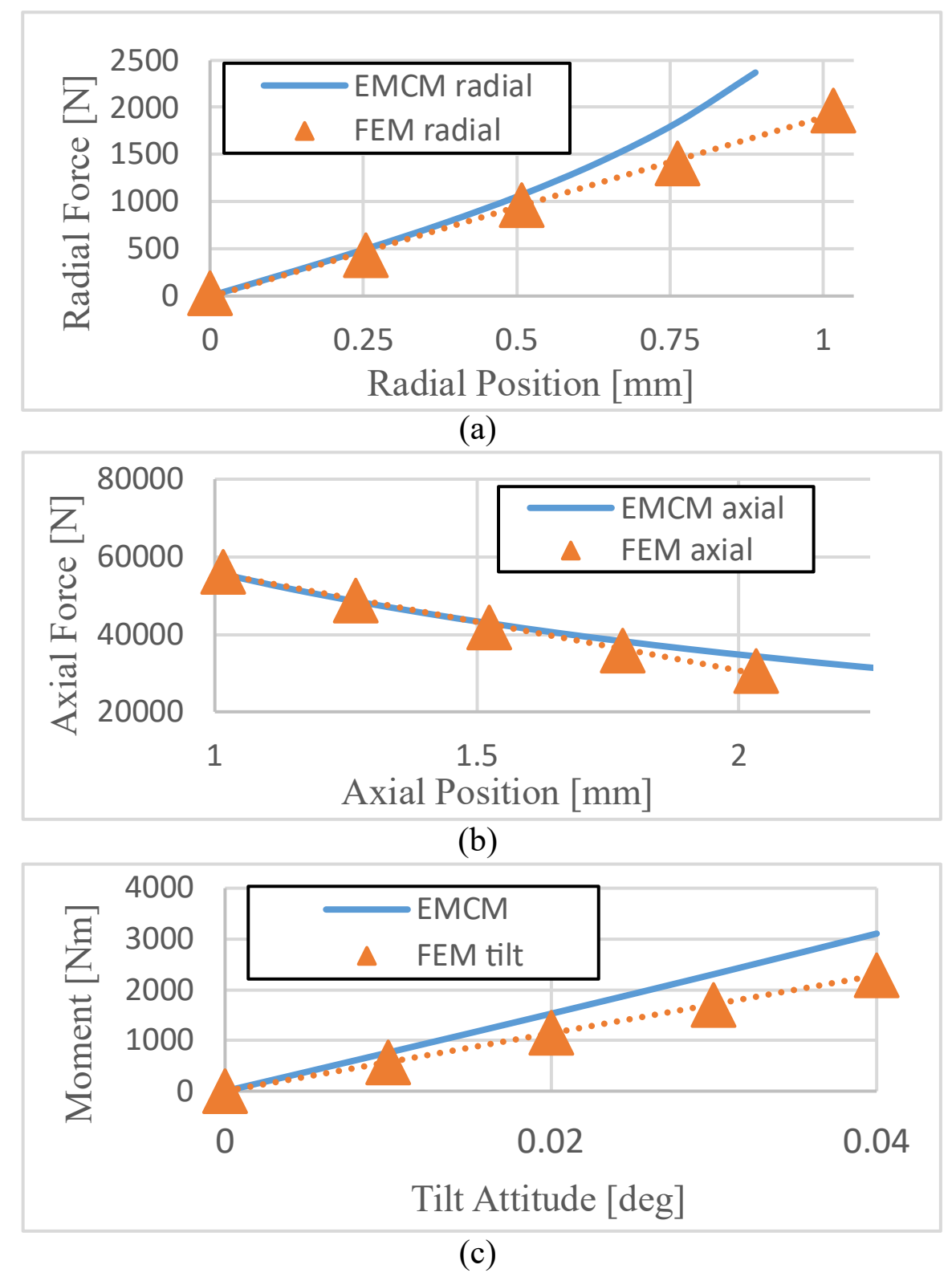

Fig. 8. (a) The radial force vs radial position. (b) The axial force vs. axial position. (c) The tilt moment vs. tilt attitude.

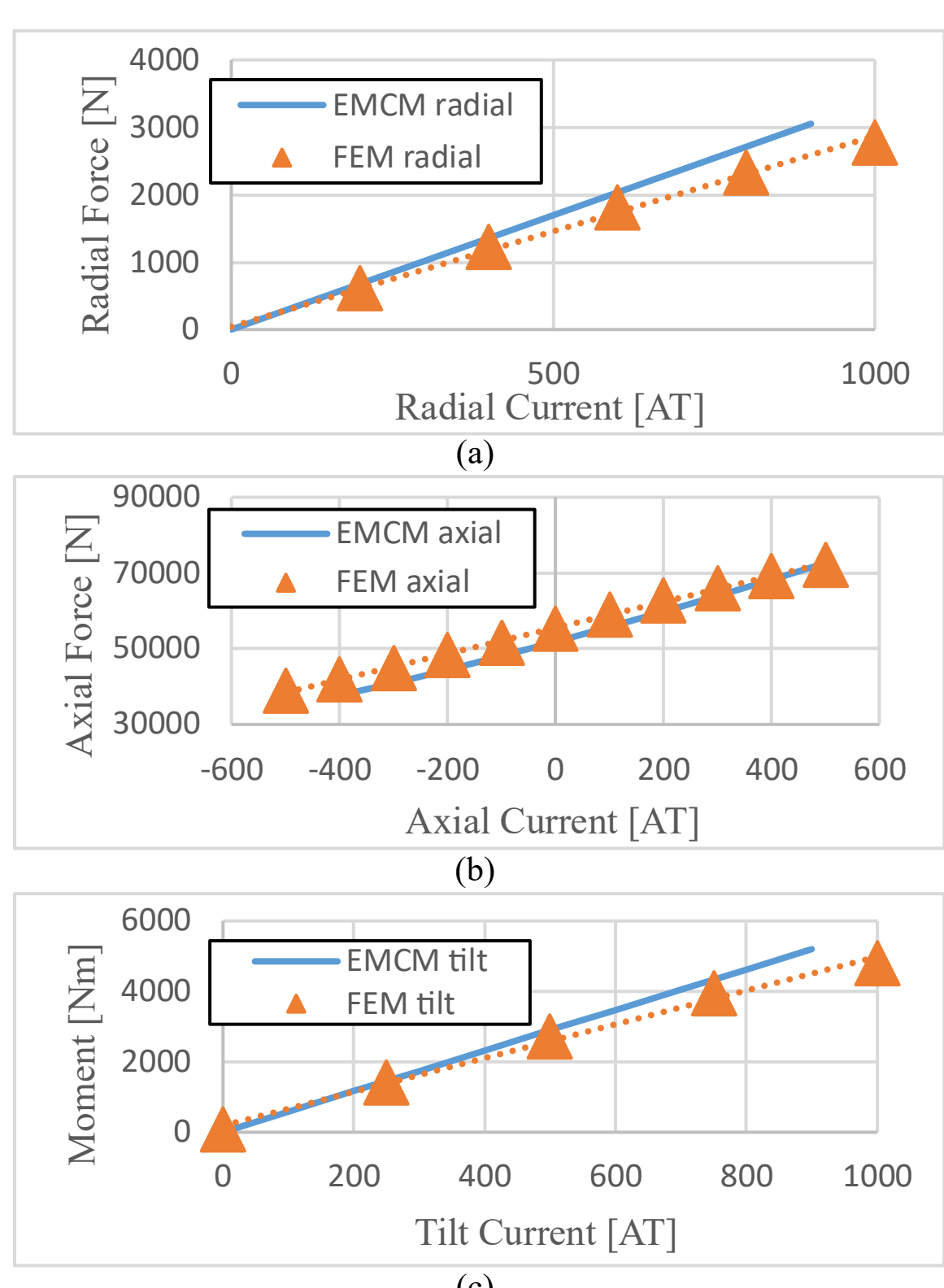

Fig. 9. (a) The radial force vs radial current. (b) The axial force vs. axial current. (c) The tilt moment vs. tilt current.

Then, the force-position, force-current, moment-position, and moment-current characteristics are analyzed by the EMCM. The coupling effects between the five axes are also investigated. Since the FEM simulation costs significant time and resources, it is mainly used for verification purposes. The FEM simulation adopts a 3D, solid mesh model with a nonlinear and adaptive solver to ensure converged results. Air gaps are modeled by several layers to improve accuracy. Stranded wire models are used for the current excitations. Over a million tetrahedral elements are used in the FEM simulation.

### A. *Force/Moment vs. Position*

The force/moment-position relationship is analyzed when a zero current is applied to the magnetic bearing. In Fig. 8, EMCM and FEM results are given for comparison. Fig. 8a shows the radial magnetic force to its radial position. The total allowable radial displacement is 1.27 mm. In the EMCM results, the radial force is linear up to 0.5 mm from the neutral position while the 3D FEM result is up to 1 mm. Unlike the EMCM, the FEM models use a nonlinear B-H curve, which captures reluctance drops in magnetic poles when the air gap increases and causes wider linear ranges. Fig. 8b shows the axial force-position relations. It shows an excellent linear relationship between axial position and force ($R^2 > 0.95$). The moment caused by the attitude change of the flywheel is depicted in Fig. 8c. Both the EMCM and FEM reveals good linearity ($R^2 > 0.95$), while the EMCM gives higher estimations. Generally, the EMCM overestimates the force/moment because it has not taken saturation and fringing effects into account.

### B. *Force/Moment vs. current*

The force/moment-current plots are acquired when the flywheel is in the equilibrium position. Fig. 9a shows the radial force vs. the current excitation, given by ampere-turns (AT). Since the radial coil is designed to have 100 turns and have a maximum excitation current of 10A, it has a maximum ampere-turn of 1000 AT. Fig. 9b shows the axial force vs. the current excitation. Here, the current is set to -500 AT to 500AT. Finally, the moment-current relation is depicted in Fig. 9c. In general, the EMCM results agree with the FEM for force/moment-current relations, with less than 20% difference.

### C. *Position Coupling effects*

Because the C5AMB has shared flux paths between the radial, PM, and axial poles, these actuators' coupling effects are investigated. For the coupling effects between the radial actuators, the y-axis displacement creates only a negligible radial force in the x-direction. However, it also affects radial reluctance and bias fluxes, which subsequently affects the x-axis force-position relationship. In detail, when y motion is significant in either direction, the x-axis will have a larger position stiffness and become more nonlinear. Fig. 10a depicts the coupling effects between x and y motions. The flywheel's tilting attitudes create a minor change in the equivalent axial reluctance that impacts the axial force. As depicted in Fig. 10b, a significant tilting attitude creates extra axial force. When $\theta_x$ and $\theta_y$ both are tilted at 0.04deg, the axial force is 6% larger than its nominal value. Nevertheless, when the tilt angles are smaller, such as between 0 and 0.02deg, the axial force is almost unaffected (less than 1%). As shown in Fig. 10 c, there is no notable moment coupling effect between tilt motions.

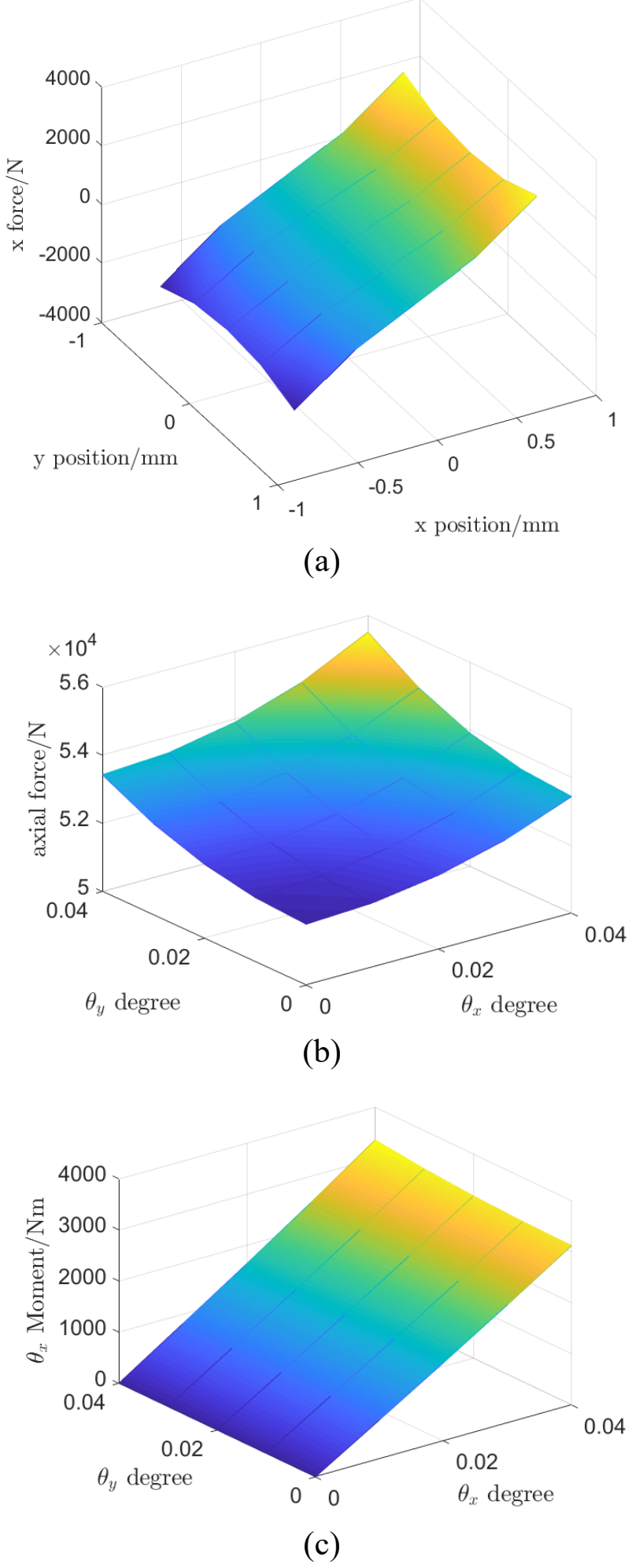

Fig. 10 . (a) X-axis forces caused by radial positions. (b) Axial force caused by attitude positions. (c) X-axis moments caused the flywheel's attitude.

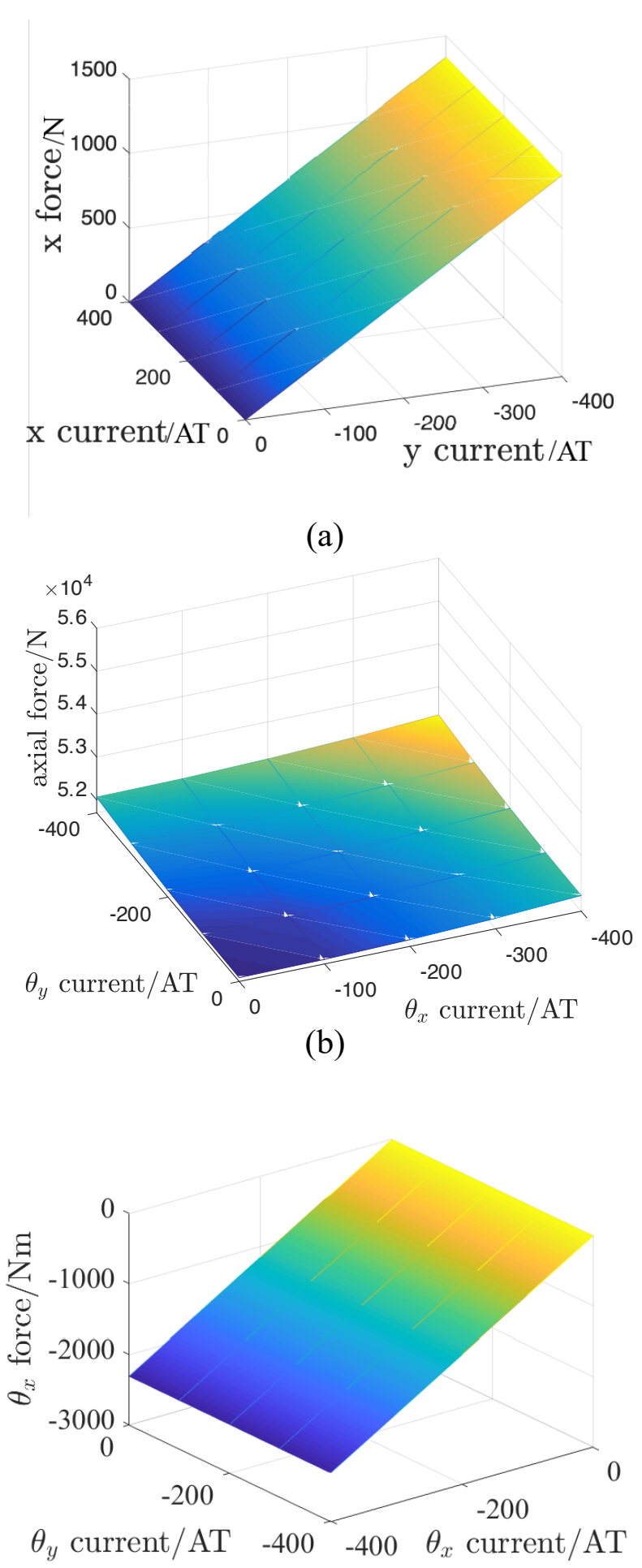

Fig. 11. (a) X-axis forces caused by radial currents. (b) Axial forces caused by tilt currents. (c) X-axis moments caused by tilt currents.

### *D. Current Coupling effects*

Coupling effects between current excitations are summarized in Fig. 11. As depicted in Fig. 11a, the radial poles show minimal coupling effects because radial control currents are applied in pairs with reversed directions to generate a net-zero flux contribution to the rest poles. Like the radial poles, tilting control currents show little to no impact on the axial force. For example, as depicted in Fig. 11b, a -400 AT tilting currents at $\theta_x$ and $\theta_y$ only cause the axial force to increase by less than 1%. Finally, there is no notable coupling effect between the tilting current excitations, as shown in Fig. 11c.

## V. Assembly and Test

In this section, the current and position stiffnesses are measured experimentally. Then, the measurements are compared to the calculated results to validate the equivalent model and the proposed design methodology. The measured bearing coefficients are also used for the flywheel's high-speed control and simulation [15]. Along with the SHFES, the C5AMB is fabricated, assembled, and tested at an off-campus facility. The magnetic bearing control algorithm is implemented in a real-time micro-controller that drives five independent power amplifiers. The controller takes feedback signals from several proximity probes that monitor the flywheel's radial, axial, and tilting motions. The feedback control algorithm for each channel includes a proportional and derivative controller, several phase-lead and lag compensators, and various band-rejection filters.

The C5AMB successfully levitated the SHFES. More details of the levitation control and testing process are described in [14], [15]. At least two notch filters are used to handle the runout effects, which will otherwise cause an excessive burden on the power amplifiers and de-levitate the rotor. When the flywheel is levitated, the current consumption is fine-tuned to less than 0.5 Ampere for radial, tilting, and axial PAs. The amplifier outputs for all actuators are controlled under 10V pk-pk. Fig. 12 depicts the complete C5AMB-SHFES assembly during testing. The flywheel is kept at its magnetic neutral position by feedback control to eliminate any magnetic force caused by displacements for measuring the current stiffness. As shown in Fig. 13, a load cell with flexible support is used to apply an external force to the flywheel and record the reaction force. In the meantime, the lag compensator is switched on to force the flywheel to stay at the neutral position. The current stiffness $K_i$ is characterized as the ratio of the applied external force to the consumed current, given by

$$K_i = f_{re}/\Delta i \tag{17}$$

where $f_{re}$ is the recorded reaction force, $\Delta i$ is the current measurement. For measuring position stiffness, the flywheel must deviate from its magnetic neutral position. For this purpose, the load cell support is replaced with a rigid structure and, the lag compensators are switched off. A new set of data is recorded for the reaction force ($f_{re}$), current ($\Delta i$), and position ($\Delta p$). In this case, the position deviation is not zero ($\Delta p \neq 0$). From $K_i$ , the position stiffness $K_p$ is derived by

$$K_p = (f_{re} - K_i \Delta i)/\Delta p. \tag{18}$$

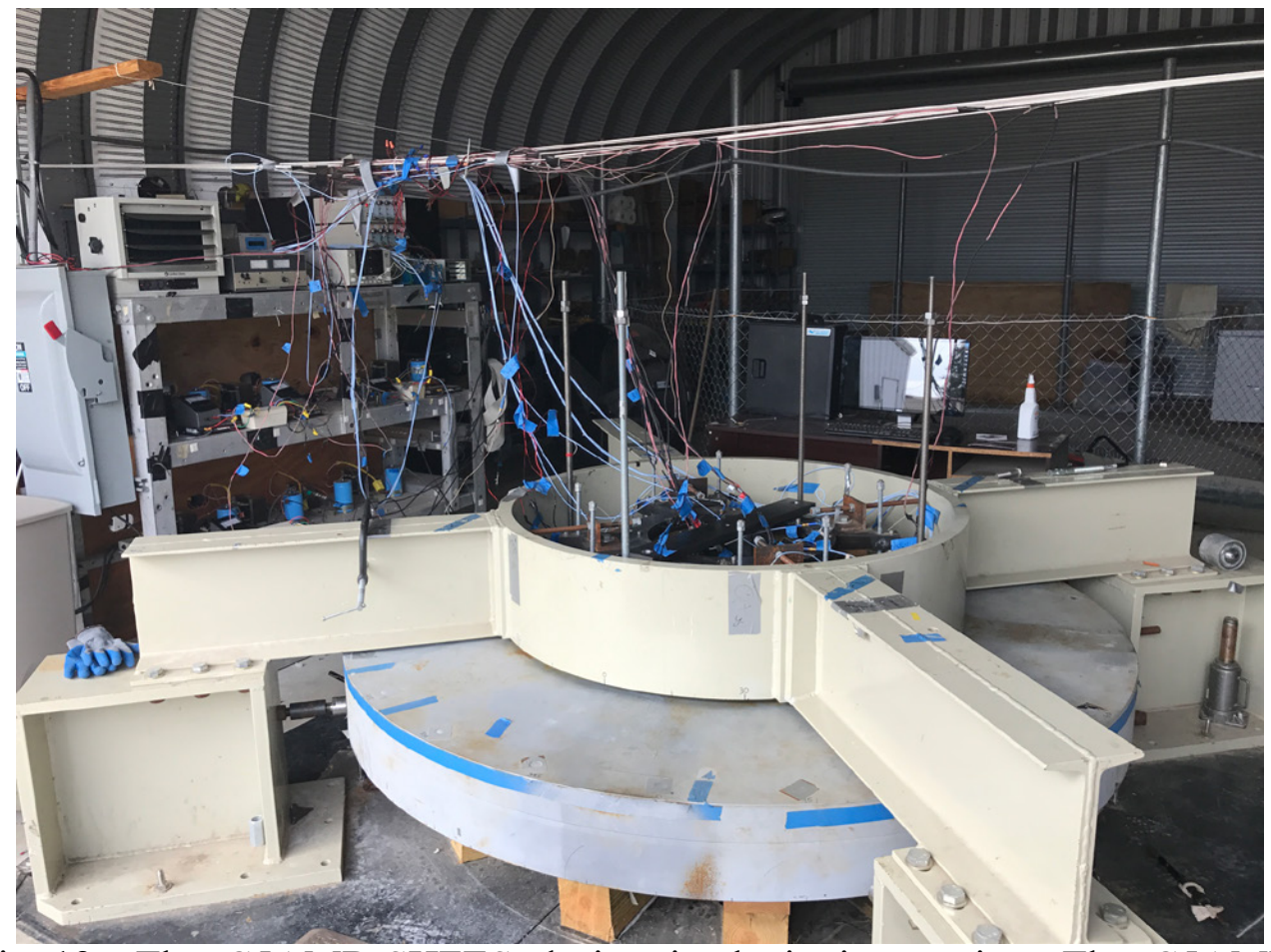

Fig. 12. The C5AMB-SHFES during its levitation testing. The C5AMB provides 53400 N to support the weight of the flywheel.

Similar procedures are carried out in the axial, tilting, and radial axis. In Fig. 14, the measured force/moment-position/attitude data is compared with FEM and EMCM simulation results. The measurement range is limited by the load cell's capacity and the power amplifier's current rating. Consequently, axial position stiffness is only evaluated between .02 mm and .08 mm of increment to the nominal air gap. A further increase of the external force will cause the axial coil to overheat. Nevertheless, the result is still helpful for modeling and controller design since the flywheel will be controlled closely to equilibrium position during normal operations. The measured force/moment-current data is depicted in Fig. 15. They are also in good agreement with either FEM or EMCM estimation. From the testing data, linear regressions are used to calculate the stiffness coefficients. The coefficients of determination ($R^2$) for all the data points are above 0.9, showing that the magnetic bearing's coefficients follow a linear pattern crucial to the flywheel's high-speed control [15].

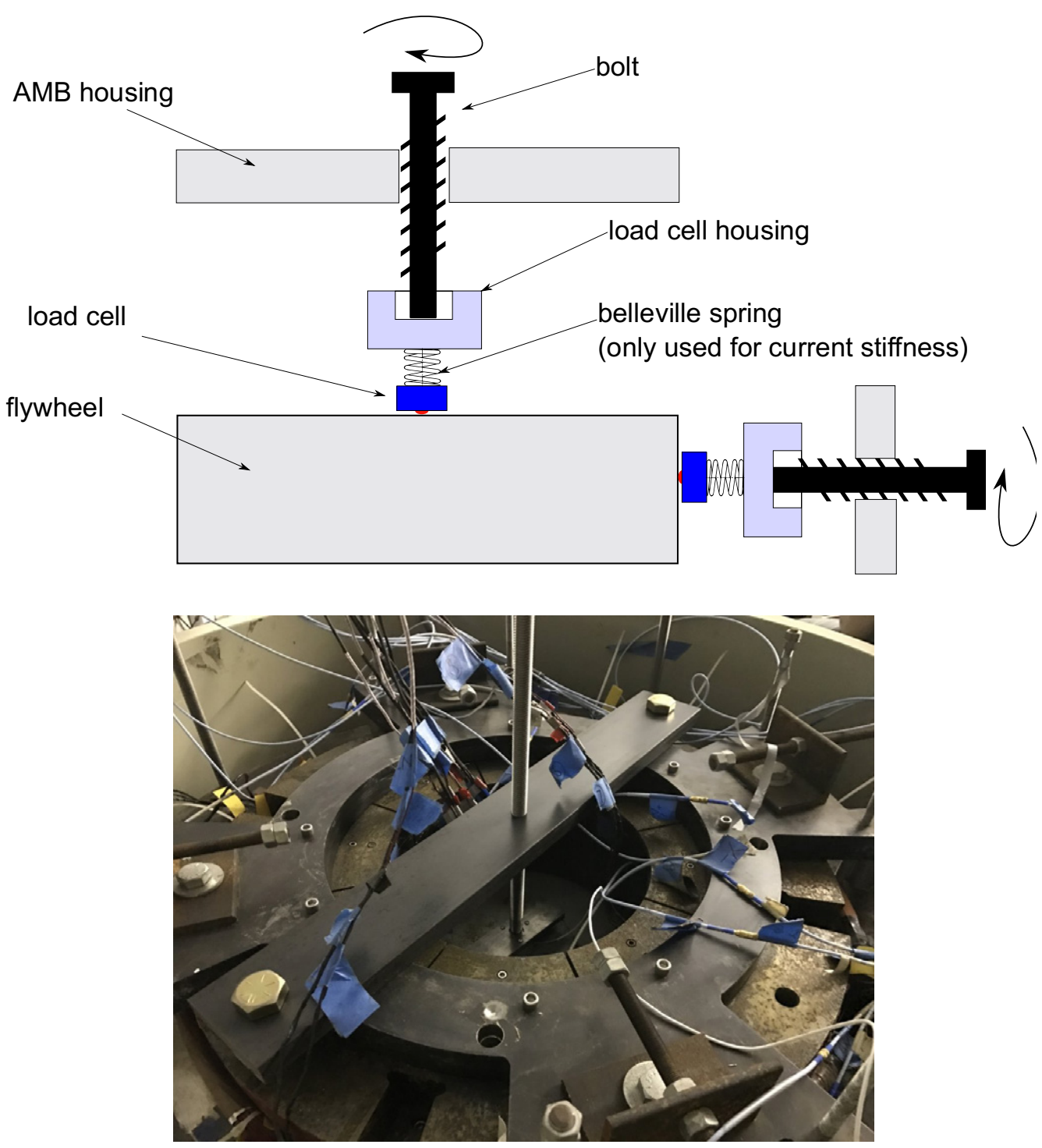

Fig. 13. The test configuration for bearing coefficients.

The position and current stiffness are summarized in Table II. In general, the EMCM has higher estimations because it uses linear magnetic permeabilities. Instead, nonlinear BH curves are used in the FEM simulations. The difference between measurements and the FEM for axial position stiffness is only about 4%, while the EMCM result has a somewhat 10% higher estimation. Similarly, titling position stiffness of FEM and measured data are closer (within 20%) than the EMCM estimation. For the radial position stiffness, FEM and EMCM estimations are within 10% to 20% of the measured results. Relatively speaking, tilting and radial position stiffness have more significant discrepancies. This can be explained by the imperfect surface flatness and radial roundness, which is expected for such a large device. Since the air gaps are small, these mechanical imperfections will have noticeable impacts. In general, current stiffnesses are more consistent since they are less affected by machining errors. For the axial current stiffness, the difference between FEM and test is within 7%. EMCM again has a higher difference of 15%. The EMCM is 20% higher for radial current stiffness, which can be attributed to linear permeability. The FEM estimation is within 10%. The tilting pole's current stiffness has the smallest difference of less than 6% between the measurements and both simulations.

TABLE II
CURRENT AND POSITION STIFFNESS

| Stiffness | | Axial | Tilting | Radial |
|---|---|---|---|---|
| *current* | *EMCM* | 4114 N/A | 576 Nm/A | 390 N/A |
| | *FEM* | 3480 N/A | 533 Nm/A | 308 N/A |
| | *test* | 3714 N/A | 546 Nm/A | 343 N/A |
| *position* | *EMCM* | -28211 N/mm | -77134 Nm/deg | -2023 N/mm |
| | *FEM* | - 24988 N/mm | -57200 Nm/deg | -1539 N/mm |
| | *test* | -25987 N/mm | -47762 Nm/deg | -1858 N/mm |

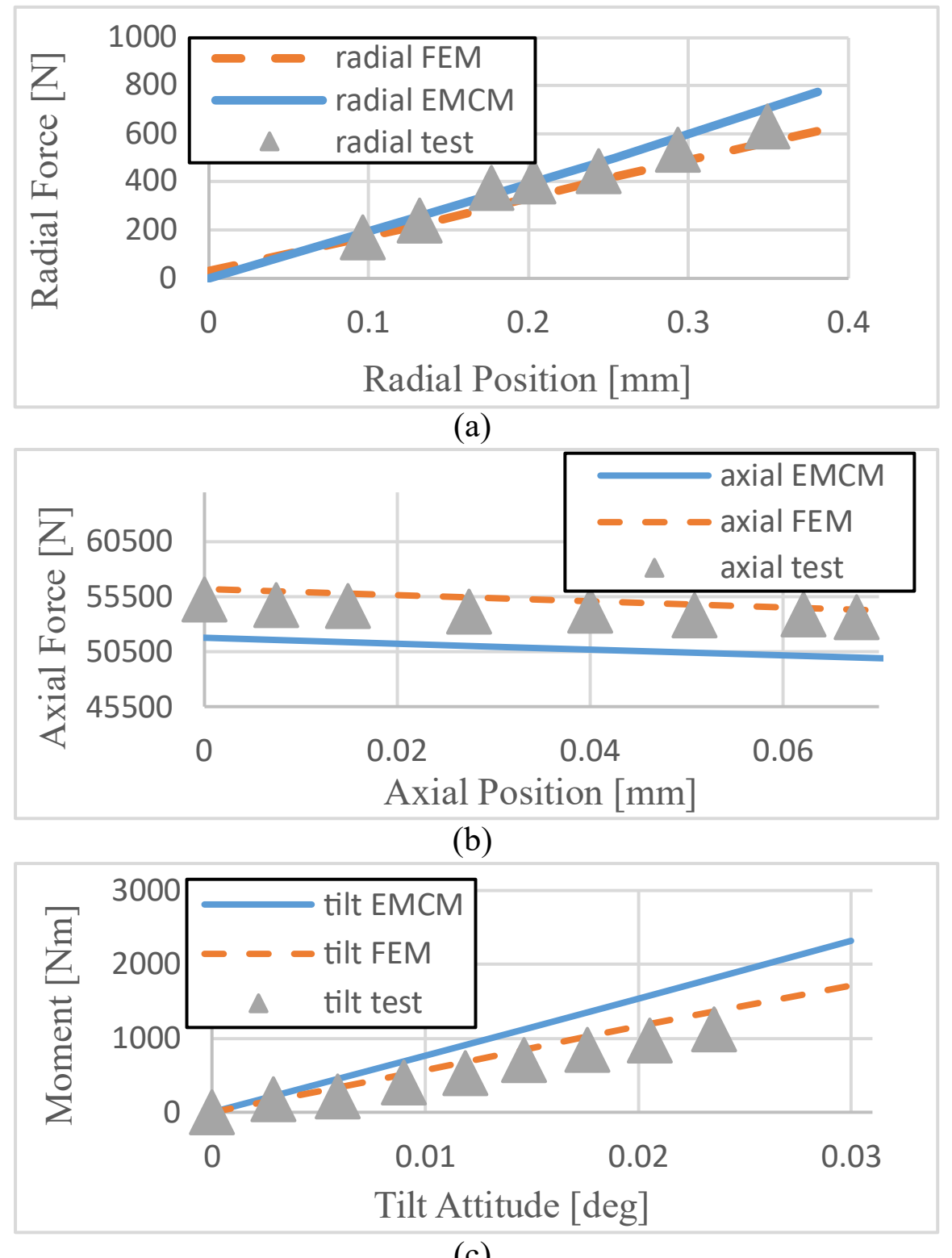

Fig. 14. (a) The measured, FEM and EMCM radial force vs. radial position. (b) The measured, FEM and EMCM axial force vs. axial position. (c) The measured, FEM and EMCM moment vs. tilt attitude.

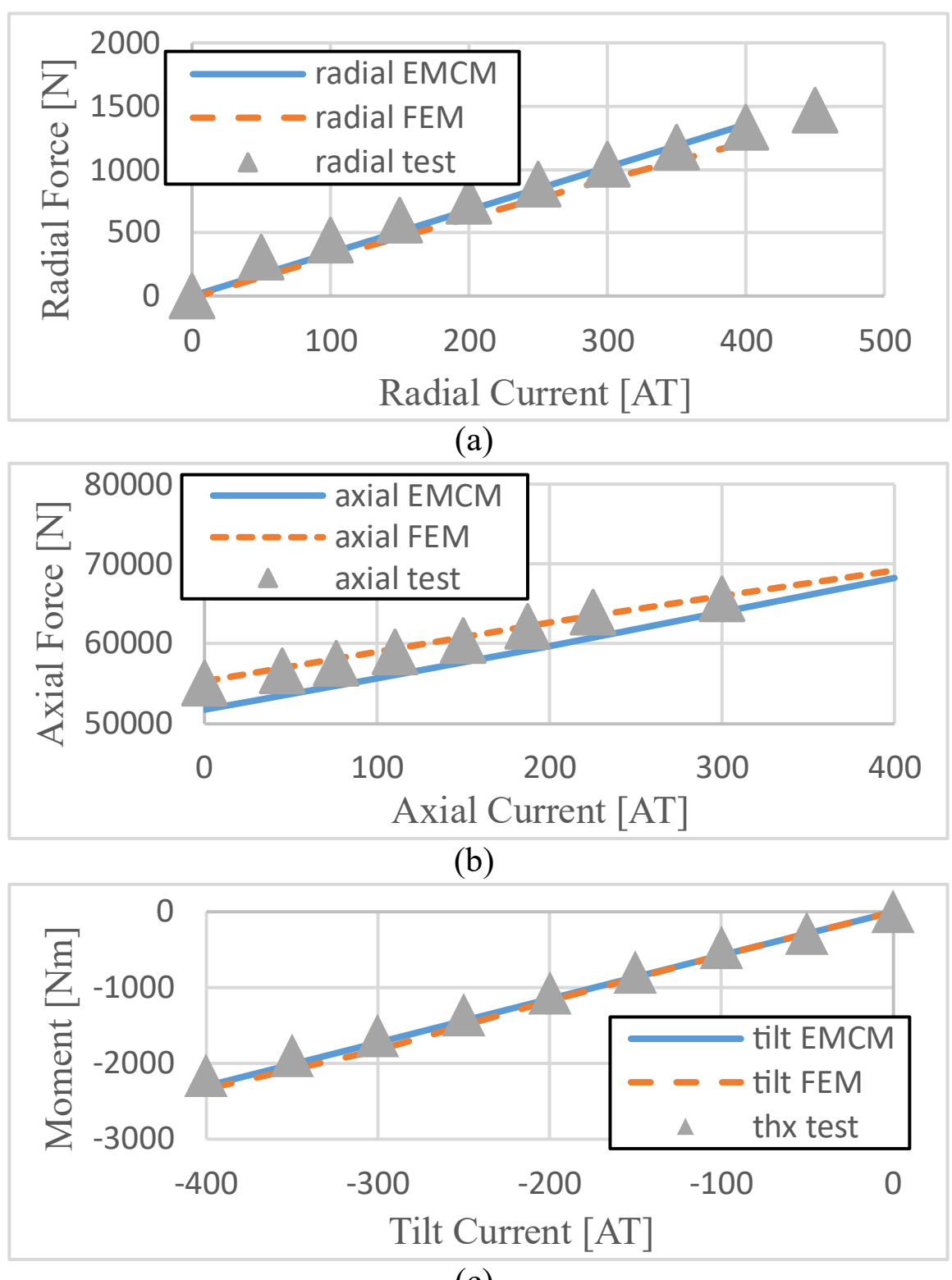

Fig. 15. (a) The measured, FEM and EMCM radial force vs. radial current. (b) The measured, FEM and EMCM axial force vs. axial current. (c) The measured, FEM and EMCM moment vs. tilt current.

## VI. Conclusion and Future Work

This paper presents a novel combination 5-DOF magnetic bearing that is highly integrated into a shaft-less energy storage flywheel. The proposed magnetic bearing is a crucial component for the flywheel to achieve double energy density. The novel design demonstrates that it is possible to condense the conventional magnetic bearing system, including several distributed units, to a single combinational device. The C5AMB's configurations and working principles are introduced first. The design methodology is then presented. Firstly, a 3D equivalent circuit model is developed to investigate its characteristics, including bias flux densities, current and position stiffness, and coupling effects. Then, FEM is used to validate the current and position stiffness to ensure good linearities and sufficient load capacities. Experimental results show that the magnetic bearing can provide stable levitation for the 5540-kg flywheel with minimal current consumptions. The measured current and position stiffnesses also show good agreement with the simulation results. In conclusion, the proposed methodology is proven to be efficient in designing large-scale, integrated magnetic bearing systems.

Some design limitations and future works are summarized as follows. The proposed AMB uses solid, non-silicon steel. For conventional AMBs, silicon steel and lamination are often used to suppress the eddy current effects to improve the frequency responses. Compared to laminated designs, the solid-core design has lowered cost but could induce more potent eddy-current effects [26], which will impact the control system's performance [27]. To verify its feasibility, we have studied high-speed simulations with the measured AMB bandwidth [15]. One of the future tasks is to create a cost-effective design to mitigate the solid-core's eddy-current effects. Also, radial control of the C5AMB relies on accurate roundness of the radial path, which is difficult to achieve for such a large device. A smaller floor-print design will improve machining accuracy and make the SHFES more applicable to other areas such as uninterruptible power supply.

## Acknowledgment

We are thankful to Erwin Thomas for his work in system assembly.

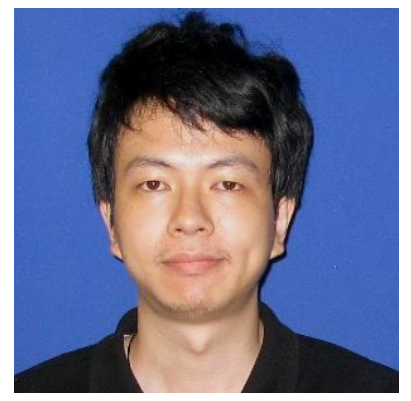

**Xiaojun Li** received his Ph.D. in engineering from Texas A&M University, College Station, TX, USA. He is currently the R&D manager at Gotion Inc, located in California, USA, where he is responsible for control and machine learning algorithms for EV and ESS batteries. Dr. Li has over 8 years of R&D experience in energy storage systems, ranging from lithium-ion batteries to energy storage flywheels.

His research interests include energy storage systems, control, and artificial intelligence. He has authored and co-authored more than 10 technical papers and has 5 issued and pending patents.

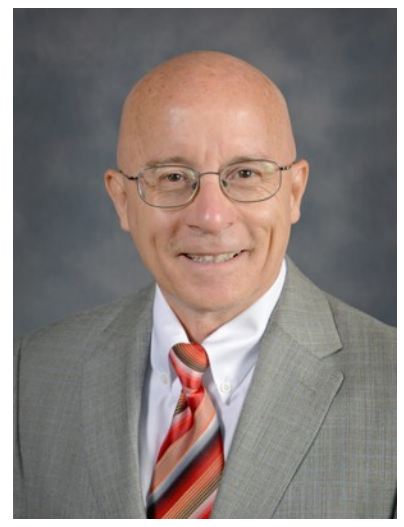

**Alan Palazzolo** received his MS and Ph.D. degrees in Mechanical Engineering from the Univ. of Virginia.

Prof. Palazzolo is an endowed professor who has received best paper awards from the ASME J. of Tribology, a lifetime achievement award in magnetic bearings, and an R&D 100 award for ultra-high-temperature magnetic bearings. Some of his current or recent research areas include centrifuge enabled desalination, flash heated direct fuel injectors, heart pumps, rotating machinery vibrations, and VFD motor-related vibration modeling. He has published 110+ archival journal articles and is the author of the textbook "Vibration Theory and Applications with Finite Elements and Active Vibration Control."

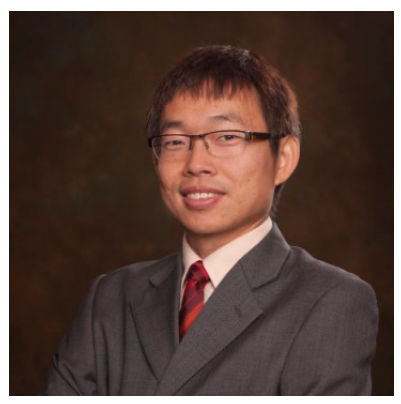

**Zhiyang Wang** received his Ph.D. in Mechanical Engineering from Texas A&M University, College Station, TX, USA.

Dr. Wang is currently the director of Research and Development in Vycon Energy, located in California, USA. His previous jobs include principal magnetic bearing engineer in Calnetix/Upwing, lead motor engineer in FMC technologies, and senior magnetic bearing engineer in GE Energy. He holds multiple US and Chinese patents on magnetic bearings, flywheels, and turbomachinery.